\documentclass[12pt]{article}
\usepackage[top=30truemm,bottom=30truemm,left=25truemm,right=25truemm]{geometry}
\usepackage{amsfonts,amsmath,amssymb,epsf}
\usepackage{amsmath,braket}
\usepackage{here}
\usepackage{physics}
\usepackage{graphicx,hyperref,color}
\usepackage[dvipsnames]{xcolor}
\usepackage{cite}
\usepackage{url}
\usepackage{comment}
\usepackage{appendix}
\usepackage{chngcntr}
\usepackage{etoolbox}
\usepackage{lipsum}

\AtBeginEnvironment{subappendices}{%
\addcontentsline{toc}{section}{Appendices}
\counterwithin{figure}{subsection}
\counterwithin{table}{subsection}
}
\hypersetup{
    unicode=false,          
    pdftoolbar=true,        
    pdfmenubar=true,        
    linktocpage=true,       
    pdffitwindow=false,     
    pdfstartview={FitH},    
    pdfnewwindow=true,      
    colorlinks=true,       
    linkcolor=blue,          
    citecolor=blue,        
    filecolor=blue,      
    urlcolor=blue           
}

\numberwithin{equation}{section}									

\newcommand{\be}{\begin{equation}}
\newcommand{\ba}{\begin{eqnarray}}
\newcommand{\ea}{\end{eqnarray}}
\newcommand{\ee}{\end{equation}}

\newcommand{\s}{\sqrt}
\newcommand{\vp}{\varphi}

\newcommand{\ti}{\tilde}
\newcommand{\ap}{\alpha}

\newcommand{\ddd}{\cdot\cdot\cdot}
\newcommand{\no}{\nonumber \\}
\newcommand{\la}{\langle}
\newcommand{\lb}{\rangle}
\newcommand{\bea}{\begin{eqnarray}}
\newcommand{\eea}{\end{eqnarray}}
\newcommand{\bes}{\begin{equation*}}
\newcommand{\beas}{\begin{eqnarray*}}
\newcommand{\eeas}{\end{eqnarray*}}
\newcommand{\bas}{\begin{array*}}
\newcommand{\eas}{\end{array*}}
\newcommand{\ees}{\end{equation*}}

\newcommand{\ep}{\epsilon}

\newcommand{\ov}{\overline}

\def\RR{\mathbb{R}}
\def\CC{\mathbb{C}}
\def\ZZ{\mathbb{Z}}

\let\a=\alpha \let\b=\beta   \let\e=\epsilon   \let\h=\eta    
\let\t=\tau   \let\vp=\varphi   \let\z=\zeta
 \let\D=\Delta

\newcommand{\Poincare}{Poincar\'{e} }

\newcommand{\corr}[1]{\left\langle #1 \right\rangle}

\usepackage{tikz}
\usetikzlibrary{arrows,arrows.meta,intersections, calc,positioning,decorations.pathreplacing,decorations.pathmorphing,shapes}
\usetikzlibrary{patterns}
\usetikzlibrary{decorations.markings}
\usetikzlibrary{knots}

\usepackage{subcaption}

\begin{document}

\begin{titlepage}
\thispagestyle{empty}

\vspace*{-2cm}
\begin{flushright}
YITP-26-17
\\
\end{flushright}

\bigskip

\begin{center}
\noindent{\bf \Large {CFT derivation of entanglement phase transition\par in pseudo entropy}}\\
\vspace{1.2cm}

Hiroki Kanda$^a$,
Tadashi Takayanagi$^{a,b}$
and Zixia Wei$^{c,a}$
\vspace{1cm}\\

{\it $^a$Center for Gravitational Physics and Quantum Information,\\
Yukawa Institute for Theoretical Physics, Kyoto University, \\
Kitashirakawa Oiwakecho, Sakyo-ku, Kyoto 606-8502, Japan}\\
\vspace{1.5mm}
{\it $^b$Inamori Research Institute for Science,\\620 Suiginya-cho, Shimogyo-ku,
Kyoto 600-8411, Japan}\\
{\it $^c$ Society of Fellows, Harvard University, \\Cambridge, MA 02138, USA}\\
\bigskip \bigskip
\vskip 1em
\end{center}

\begin{abstract}

In this paper, we discuss the entanglement phase transition of pseudo entropy in CFTs. We focus on the case where the in-state and the out-state are different boundary states related by boundary condition changing operators. We compute the pseudo entropy with BCFT methods and find a phase transition with respect to the conformal weight of the boundary condition changing operators. For holographic CFTs, we confirm that the CFT results match that evaluated in AdS.
\end{abstract}

\end{titlepage}

\newpage

\tableofcontents

\newpage

\section{Introduction}

The entanglement entropy has played major roles as a fundamental order parameter in quantum many-body systems \cite{Vidal:2002rm,Calabrese:2004eu,Kitaev:2005dm,Levin:2006zz} and moreover it is a useful quantity which extracts hidden geometrical structures inside them. Indeed the holography, especially the AdS/CFT \cite{Maldacena:1997re}
makes a definite meaning to this geometry, namely its holographic spacetime, as implied by the holographic calculation of entanglement entropy \cite{Ryu:2006bv,Ryu:2006ef,Hubeny:2007xt}.

Recently, a generalization of entanglement entropy to the case where we assume a post-section process was found in \cite{Nakata:2021ubr} and is called pseudo entropy (see also \cite{Mollabashi:2020yie,Mollabashi:2021xsd,Miyaji:2021lcq,Akal:2021dqt,Goto:2021kln,Mukherjee:2022jac,Guo:2022sfl,Ishiyama:2022odv,Miyaji:2022dna,Bhattacharya:2022wlp,Guo:2022jzs,Parzygnat:2022pax,He:2023eap,Parzygnat:2023avh,Kanda:2023zse,Kanda:2023jyi,Guo:2023aio,Balasubramanian:2023xyd,Kawamoto:2023ade,Wei:2024zez,He:2024jog,Hao:2024nhd,Caputa:2024gve,Capone:2024oim,Yan:2024rcl,Shimizu:2025kse,Milekhin:2025ycm,Cerezo-Roquebrun:2025uol,Chu:2025sjv,Nunez:2025ppd,Roychowdhury:2025ebs,Kawamoto:2025oko,Anegawa:2025tio,Nunez:2025puk,Chen:2025ibe,Hao:2025ocu,Guo:2025dtq,Carignano:2025srt,Fareghbal:2025ljs,Fujiki:2025rtx,Fukusumi:2025xrj,Li:2025tud,Dulac:2025owj,Anastasiou:2025rvz,Das:2025fcd,Harper:2025lav,Caputa:2025ugm,Das:2026ifj,Fareghbal:2026vbl} for a partial list of more developments). This is defined as follows. Consider the initial state $|\psi \lb$ and perform a post-selection to another state $|\vp\lb$, which has a non-zero overlap $\la\vp|\psi\lb\neq 0$. Then we introduce the transition matrix defined as $\frac{|\psi\lb\la\vp|}{\la\vp|\psi\lb}$. Assuming the decomposition of Hilbert space $H_{\rm tot}=H_A\otimes H_B$, we can introduce the reduced transition matrix 
\ba
\tau_A=\mbox{Tr}_B\left[\frac{|\psi\lb\la\vp|}{\la\vp|\psi\lb}\right].
\ea
The pseudo entropy is defined by 
\ba
S_A=-\mbox{Tr}\left[\tau_A\log\tau_A\right]. \label{pe}
\ea
When the initial state and the final state coincide $|\phi\lb=|\vp\lb$, $S_A$ becomes identical to the entanglement entropy. 

Since $\tau_A$ is not Hermitian, this quantity (\ref{pe}) is complex-valued in general. However, in a special class of states, this quantity takes positive real values and we can interpret the pseudo entropy as the averaged number of Bell pairs which appear as the intermediate state of the post-selection process \cite{Nakata:2021ubr}. Moreover, if we consider a Euclidean path integral with real-valued but Euclidean time-dependent sources, the standard replica calculation \cite{Calabrese:2004eu} leads to a positive-valued pseudo entropy. In holographic CFTs, this pseudo entropy can be computed as the minimal area \cite{Nakata:2021ubr} in the same way as the holographic entanglement entropy. 
When we consider the Lorentzian time evolutions of the pseudo entropy, a minimal surface should be replaced with an extremal surface which typically extends in the directions of complexified coordinates \cite{Mollabashi:2020yie}.
Such complex extremal surfaces are also common in the time-like entanglement entropy \cite{Doi:2022iyj,Doi:2023zaf}
(see also \cite{Narayan:2022afv,Narayan:2023ebn,Narayan:2023zen}) as pointed out in \cite{Heller:2024whi,Heller:2025kvp} and the pseudo entropy in dS/CFT \cite{Fujiki:2025rtx}.

One of the most basic dynamical properties of entanglement entropy in CFTs is that it grows linearly under the time evolution if we consider the time evolution of a product state, as highlighted by the quantum quench process \cite{Calabrese:2005in}. This growth describes the thermalization procedure in CFTs, induced by the entangled pairs propagating at the speed of light. A convenient description of the product state, which does not have real space entanglement, can be found by considering a boundary conformal field theory (BCFT), namely a conformal field theory (CFT) which lives on a manifold with conformal boundaries. In this description, the product state is described by a boundary state (Cardy state) $|B\lb$ \cite{Cardy:1984bb,Cardy:1989ir,Cardy:2004hm,Miyaji:2014mca}. Each boundary condition corresponds to its own boundary state, which can be written as $|B_a\lb$, where $a$ denote the label of boundary condition.

Thus, it is a natural and intriguing question to ask how the pseudo entropy evolves under the time evolution of a product state in the presence of a post-selection to another quantum state. Recently, in the papers \cite{Kanda:2023zse,Kanda:2023jyi}, this question was studied by employing a bottom up construction of gravity dual of a boundary conformal field theory, so called AdS/BCFT \cite{Takayanagi:2011zk,Fujita:2011fp}. In this model, an initial and final state are given by two different regularized boundary states 
\be
|\psi\lb=e^{-\frac{\beta}{4}H}|B_a\lb,\ \ \ \  |\vp\lb=e^{-\frac{\beta}{4}H}|B_b\lb,
\label{BSca}
\ee
respectively. These two boundary states are related to each other by an exactly marginal perturbation and the values $a$ and
$b$ describe those of the moduli. When $a-b=0$, the two states are identical. The pseudo entropy (\ref{pe}) for these states turned out to experience an interesting phase transition. When the difference is small such that $|a-b|<\gamma_*$ for a certain positive constant $\gamma_*$, then the pseudo entropy shows a linear growth under the time evolution: $S_A\propto t$. In particular, when $a=b$, this matches with the gravity dual of quantum quench \cite{Hartman:2013qma}. However, when $|a-b|>\gamma_*$. the pseudo entropy become time-independent: $S_A=$const. Moreover, in two dimensional holographic CFTs, at the critical point $|a-b|=\gamma_*$, the logarithmic growth was observed 
$S_A\simeq \frac{c}{3} \log t$ \cite{Kanda:2023jyi}.

This phase transition of pseudo entropy looks very analogous to the entanglement phase transition or measurement induced phase transition \cite{Li:2018mcv,Skinner:2018tjl,Li:2019zju,Kawabata:2022biv}, which has attracted much attention in the context of quantum many-body physics. In this context, the time evolution of entanglement entropy shows a similar phase transition when we introduce a non-unitarity induced by partial projection measurements under the time evolution. When the effect of non-unitarity is small, the entanglement entropy grows linearly, while when the non-unitarity gets larger, it becomes time-independent. At the border between these two phases, there is a critical point where the entropy grows logarithmically. In the model of  \cite{Kanda:2023zse,Kanda:2023jyi}, the time evolution is unitary but a final state projection is made in the end, while in the 
model of measurement induced phase transition, the time evolution is non-unitary but there is no final state projection. If we remind that the real part of pseudo entropy measures the amount of entanglement in the intermediate states between the initial and final state, then the similarity of the results from two different models is not surprising.

Since the holographic analysis based on the AdS/BCFT \cite{Kanda:2023zse,Kanda:2023jyi} has been done purely in the gravity dual side using the effective description of localized scalar field on the end-of-the-world brane, it is not precisely clear 
what BCFT this corresponds to microscopically. This makes the prediction of the gravity results less clear. To overcome this problem, in this paper we consider a different setup of computing the time evolution of pseudo entropy 
in a holographic CFT, where both the CFT and its gravity dual are explicit, 
and confirm that the phase transition of pseudo entropy indeed occurs. For this we will start with two identical boundary states in a holographic CFT and act boundary primary operators in between them so that the two boundary states differ from each other. By inserting two twist operators as usual in the replica method \cite{Calabrese:2004eu} and applying the heavy-heavy-light-light conformal block approximation \cite{Fitzpatrick:2014vua,HHLLCorr,Fitzpatrick:2015zha}, we are able to compute the pseudo entropy when the initial and final state are given by two different boundary states. From this calculation we will observe a phase transition, which is similar to the measurement induced phase transition. As an opposite example, we will also perform a similar calculation for a massless free Dirac fermion CFT. In this case, we will find that the behavior of pseudo entropy does not change at all even if we choose different boundary states. 

The contents of this paper is as follows.
In section two, we will perform the calculations of pseudo entropy in a holographic CFT and show that it experiences the phase transition. In section three, we will compute the pseudo entropy in a free Dirac fermion.
In section four, we summarize our conclusion.
In appendix A, we present the details of the calculations of the pseudo entropy in the Dirac fermion CFT.

\section{Entanglement Pseudo Transition in Holographic CFTs}

\begin{figure}[hbpt]
\begin{minipage}[b]{0.5\hsize}
    \centering
    \includegraphics[width=0.9\linewidth]{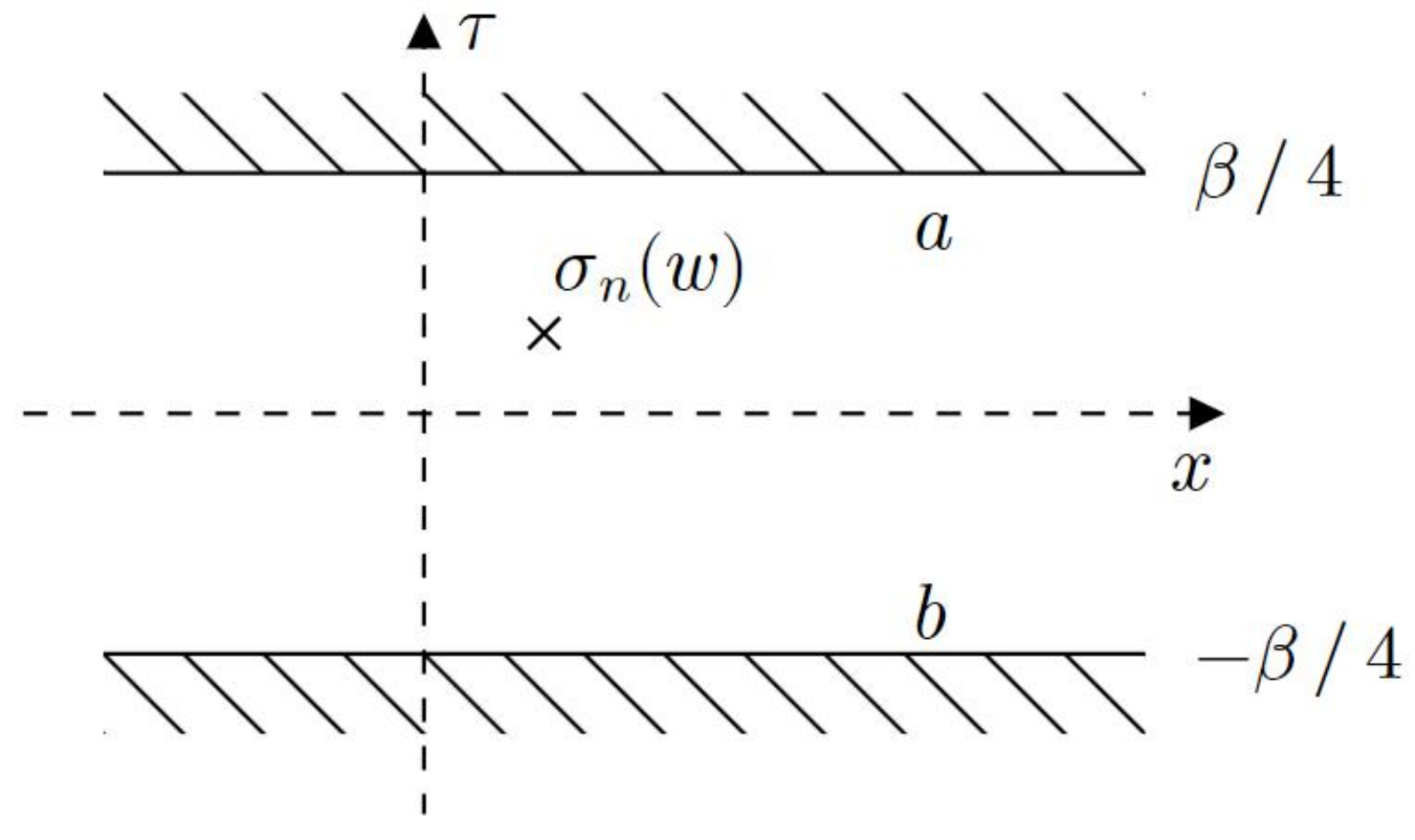}
    \subcaption{}
    \label{fig:StripWithDiffBC}
\end{minipage}
\begin{minipage}[b]{0.5\hsize}
    \centering
    \includegraphics[width=0.9\linewidth]{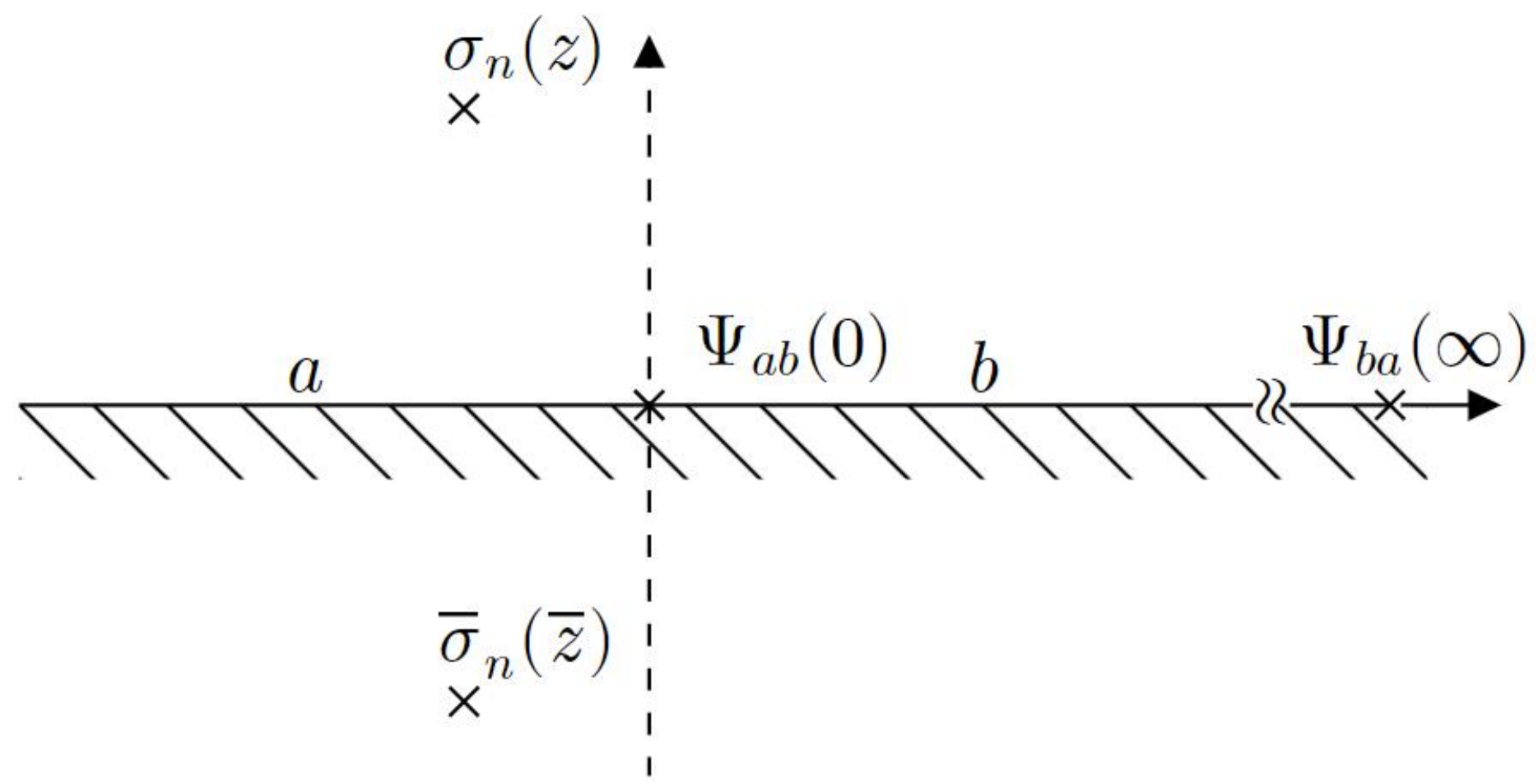}
    \subcaption{}
    \label{fig:UHPWithBCC}
\end{minipage}
\caption{(a) shows the strip geometry to calculate the pseudo entropy. $a$ and $b$ denote boundary conditions, or boundary states. (b) shows the upper half geometry after the conformal mapping and its mirror image appearing from the doubling trick.}
\end{figure}

In this section, we calculate the entanglement entropy for a half line subsystem $A: x>0$, on the strip with a width of $\b/2$, imposing different boundary conditions on each boundary.
This can be computed from the one point function of twist operator on the upper half plane, using the replica method as shown in Fig.\ref{fig:StripWithDiffBC}. Since we set the imaginary time perpendicular to strip, the analytical continuation of the entanglement entropy actually becomes pseudo entropy $S_A$ defined by (\ref{pe}) for the two quantum states (\ref{BSca}).

\subsection{Analysis in Euclidean Setup}

First, using conformal mapping $z=ie^{2\pi w/\b}$, this strip is transformed into the upper half plane, shown in Fig.\ref{fig:UHPWithBCC}. Furthermore, two boundary condition changing (b.c.c.) operators, $\Psi_{ab}$ and $\Psi_{ba}$, are inserted at $z=0,\infty$. 
{Note that different from the twist operator, which is a bulk operator (corresponding to a closed string state) in BCFT, a b.c.c. operator is a boundary operator (corresponding to an open string state).  Also that the boundary operator $\Psi_{ab}$ only admits a chiral conformal dimension $h_{\Psi}$ without an anti-chiral part.} 
{Hence, the one-point function of the twist operator $\sigma_n(w)$ becomes
\begin{align}
    \corr{\sigma_n(w)}_{ \rm{strip}} &= \left(\frac{2\pi}{\b}\abs{z}\right)^{\D_n}\corr{\Psi_{ba}(\infty)\sigma_n(z)\Psi_{ab}(0)}_{\rm{UHP}}
\end{align}
where $\sigma_n(z)$ acts as a twist operator with conformal dimension 
\begin{equation}
    h_n= \bar{h}_n =  \frac{c}{24}\left(n-\frac{1}{n}\right).
\end{equation}
To evaluate the boundary-boundary-bulk 3-point function $\corr{\Psi_{ba}(\infty)\sigma_n(z)\Psi_{ab}(0)}_{\rm{UHP}}$, it is useful to apply the mirror method (or the doubling trick) in BCFTs \cite{Lewellen:1991tb,SvRW20,Bernamonti:2024fgx}, which states that the kinematics of $\corr{\Psi_{ba}(\infty)\sigma_n(z)\Psi_{ab}(0)}_{\rm{UHP}}$ is identical to that of the following chiral 4-point function on $\mathbb{C}$,
\begin{align}
    \corr{\Psi_{ba}(\infty)\sigma_n^L(z)\tilde{\sigma}_n^L(\ov{z})\Psi_{ab}(0)}_{\CC}, \label{eq:4pfunction}
\end{align}
where $\sigma_n^L(z)$ ($\tilde{\sigma}_n^L(\ov{z})$) is the chiral part of the twist (anti-twist) operator. 
}

{Assuming the b.c.c operators are heavy, i.e. $h_{\Psi} = \mathcal{O}(c)$, the four-point function \eqref{eq:4pfunction} is able to be approximated by the dominance of the vacuum block \cite{HHLLCorr}. Due to this feature, we have the following approximation:
\ba
    \corr{\Psi_{ba}(\infty)\sigma_n(z)\Psi_{ab}(0)}_{\rm{UHP}} &\sim& \corr{\Psi_{ba}(\infty)\sigma_n^L(z)\tilde{\sigma}_n^L(\ov{z})\Psi_{ab}(0)}_{\CC}, \no
    &\simeq &z^{-2h_n}G_n(\zeta),    \label{eq:4pfunctionn}
\ea
in this special case. Here $\zeta=\frac{\bar{z}}{z}$ is the cross ratio and the conformal block function $G_n(\zeta)$ is given by 
\begin{equation}
    \log G_n(\zeta) \simeq \frac{c}{6}(1-n)\log\left[\frac{\z^{(1-\a_\Psi)/2}(1-\z^{\a_\Psi})}{i\a_\Psi}\right],
\end{equation}
when $n-1$ is infinitesimally small. 
Here we have defined 
\begin{align}
    \a_\Psi := \sqrt{1-24\frac{h_\Psi}{c}}
\end{align}
Thus we find
\ba
   \corr{\sigma_n(w)}_{\rm{strip}} &=& \left(\frac{2\pi}{\b}\right)^{2h_n}\left(\frac{\bar{z}}{z}\right)^{h_n}G_n(\z)\no
   &=&\left(\frac{2\pi}{\b}\right)^{2h_n}\left[
   \frac{1-\z^{\a_\Psi}}{i\a_\Psi
   \z^{\a_\Psi/2}}
   \right]^{-2h_n},
\ea
where $h_n\simeq \frac{c}{12}(n-1)$ in the limit $n\to 1$.

In this way, we obtain the pseudo entropy
\begin{subequations}
\begin{align}
    S_A&=\lim_{n\to 1}\frac{1}{1-n}\log \corr{\sigma_n(w)}_{\rm{strip}}\\
    &=\frac{c}{6}\log\frac{\b}{2\pi\epsilon}+\frac{c}{6}\log\left[\frac{2\sinh(\frac{\a_\Psi}{2}\log\z)}{i\a_\Psi}\right].\label{eq:entropy_loogg}
\end{align}
\end{subequations}
where $\z=-e^{4\pi i\t/\b}$ and $\ov{\z}=-e^{-4\pi i\t/\b}$. $\log \z$ has some branch.
Especially, there are two branches for $\log \zeta$:
\begin{equation}
    \log \z = 4\pi i\frac{\t}{\b} \pm \pi i, 
\end{equation}
that should be considered.}

Now we write down the final formula of $S_A$ which depends on the parameter range of $\ap_\Psi$ as follows.
If $0<\a<1$, we find
\begin{equation}
    S_A = \frac{c}{6}\log\frac{\b}{2\pi\a_\Psi \e} + \frac{c}{6}\log\left[2\sin\left(\frac{2\pi\a_\Psi}{\b}\left(\frac{\b}{4}\pm\t\right)\right)\right],\label{eq:EE_BTZ_w_branch}
\end{equation}
else if $\a=0$,
\begin{equation}\label{eq:EE_Poincare_w_branch}
    S_A = \frac{c}{6}\log\frac{\b}{2\e} + \frac{c}{6}\log\left(1\pm\frac{4\t}{\b}\right),
\end{equation}
and if $\a\in i\RR$, then
\begin{equation}
    S_A = \frac{c}{6}\log\frac{\b}{2\pi\abs{\a_\Psi} \e} + \frac{c}{6}\log\left[2\sinh\left(\frac{2\pi\abs{\a_\Psi}}{\b}\left(\frac{\b}{4}\pm\t\right)\right)\right].\label{eq:HHLL_TAdS}
\end{equation}
These are Euclidean time evolutions of the pseudo entropy. We will work out the Lorentzian evolution in the next subsection.

\subsection{Lorentzian evolution in Holographic CFT}
\subsubsection{Small deformation phase $0<h_{\Psi}<\frac{c}{24}$}
Next, consider back-reacted geometry by the b.b.c. operators of $0 < \a_\Psi < 1$. Without b.b.c. operators, the metric of dual AdS of the strip (Fig.\ref{fig:StripWithDiffBC}) can be written as \cite{Banados}
\begin{equation}
    ds^2 = \frac{dy^2 + dw d\ov{w}}{y^2} + \frac{\pi^2}{\b^2}dw^2 + \frac{\pi^2}{\b^2}d\ov{w}^2 + \frac{\pi^4}{\b^4}y^2dwd\ov{w},
\end{equation}
which is the Euclidean BTZ metric. by coordinate transformation
\begin{equation}
    z = \frac{\b^2}{\pi^2}\left(y + \frac{\b^2}{\pi^2}y^{-1}\right)^{-1},\quad w = x + i\t,\quad \ov{w} = x - i\t
\end{equation}
the metric reduces into
\begin{equation}
    ds^2 = \frac{dz^2}{z^2 h(z)} + \frac{h(z)d\t^2}{z^2} + \frac{dx^2}{z^2},\quad h(z) = 1 - \frac{z^2}{a^2},\quad a = \frac{\b}{2\pi},
\end{equation}
where $a$ is the radius of the BTZ black hole. Due to insertion of the b.c.c. operator, the dual geometry admits the deficit angle. Thus, the corresponding metric becomes
\begin{equation}
    ds^2 = \frac{dz^2}{z^2h(z)} + \frac{h(z)d\t^2}{z^2} + \frac{dx^2}{z^2},\quad a= \frac{\b}{2\pi \a_\Psi}
\end{equation}
where $\a$ controls the angular deficit. 

\begin{figure}[htbp]
\begin{minipage}[b]{0.45\hsize}
    \centering
    \includegraphics[width=0.7\linewidth]{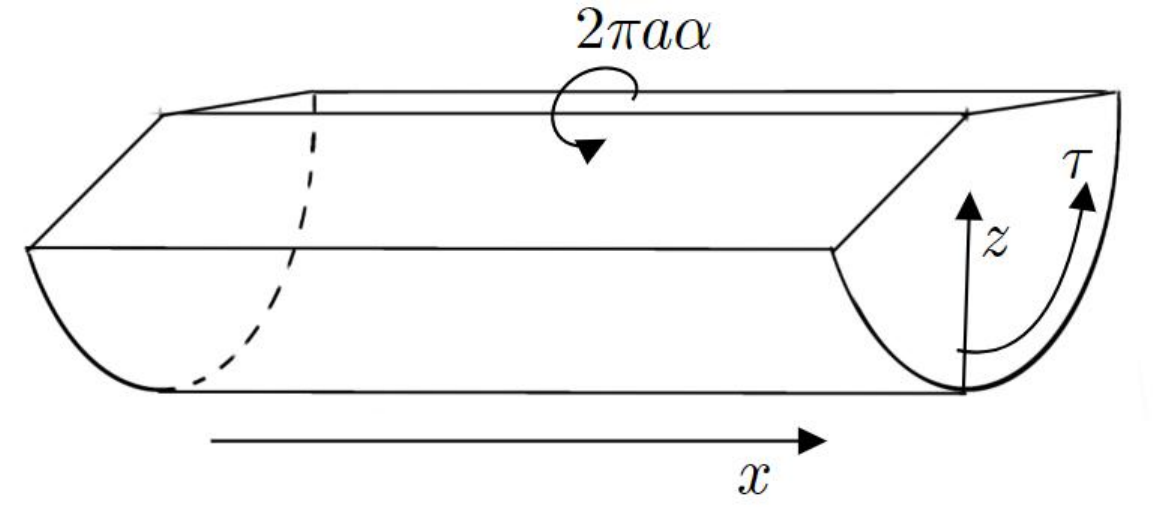}
    \caption{The Euclidean BTZ geometry. This has a conical singularity with $\tau$-direction}
    \label{fig:deficitBTZ}
\end{minipage}
\hspace{0.1\hsize}
\begin{minipage}[b]{0.45\hsize}
    \centering
\includegraphics[width=0.7\linewidth]{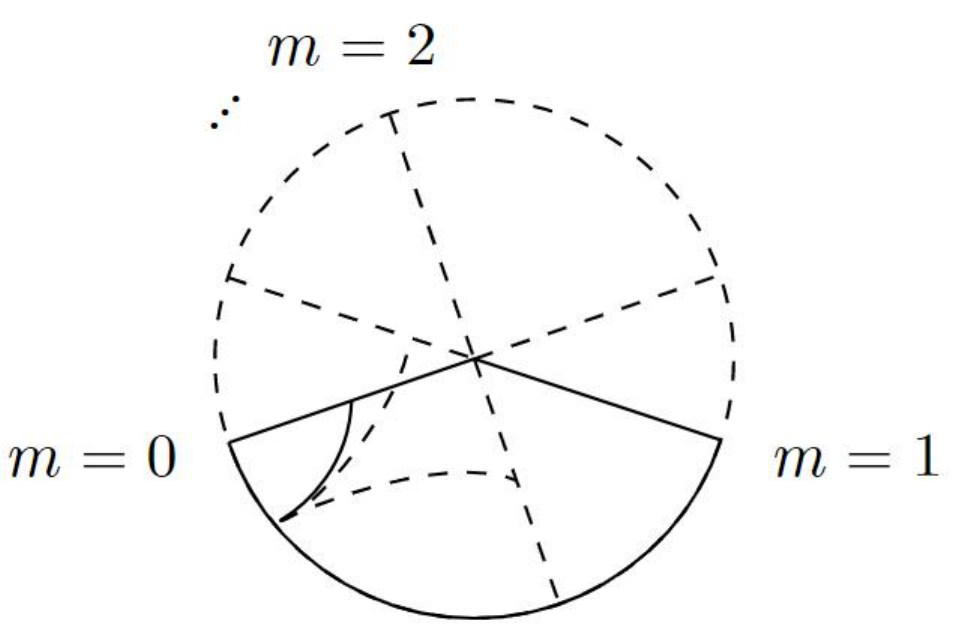}
    \caption{There are two candidates of geodesics, which are complex conjugate to each other in the Lorentzian space}
    \label{fig:unphysicalBrane}
\end{minipage}
\end{figure}

Consider the embedding AdS$_{3}$ into $\RR^{2,2}$.
\begin{equation}
    \h_{MN}=\operatorname{diag}(-1,0,0,-1), \quad \h_{MN}X^M X^N=-1
\end{equation}

In this geometry, the geodesic curve is described as
\begin{equation}
    X^{M} =\frac{\sinh(( 1-s) L)}{\sinh L} X_{0}^{M} +\frac{\sinh( sL)}{\sinh L} X_{1}^{M},\quad s\in C,\quad \cosh L=-\h_{MN}X^M_0X_1^N
\end{equation}
where $C$ is any integral curve from $0$ to $1$ that is homotopically equivalent to the segment $[0,1]$. Moreover, the geodesic length of this curve is $L$, which is independent of the choice of integral path. Here, to observe the BTZ metric, we should set
\begin{equation}
    X^0 = \frac{a}{z}\cosh\frac{x}{a},\quad X^1 = \frac{a}{z}\sinh\frac{x}{a},\quad X^2 = \sqrt{\frac{a^2}{z^2}-1}\cosh\frac{i\tau}{a},\quad X^3 = \sqrt{\frac{a^2}{z^2}-1}\sinh\frac{i\tau}{a}.
\end{equation}
Now, the Ryu-Takayanagi surface is the geodesic connecting the insertion point of the twist operator $(\epsilon,\tau,0)$ and the point on the EOW brane $(z_*, \pm\b/4,0)$. We choose $z_*$ to give the minimal geodesic length. Thus, the geodesic length is
\begin{equation}
    L \simeq \frac{a}{\epsilon }\sin\frac{\b/4-\abs{\t}}{a}
\end{equation}

Therefore we obtain the entanglement entropy in the usual manner as following
\begin{equation}
    S_A = \frac{c}{6}L \cong \frac{c}{6}\frac{a}{\e}\sin(\frac{\b/4-\abs{\t}}{a})= \frac{c}{6}\ln\frac{\b}{2\pi\a_\Psi\e}+\frac{c}{6}\log\left[ 2\sin\left(\frac{\pi}{2}\a_\Psi-\frac{2\pi\a_\Psi}{\beta}\abs{\t} \right)\right].
\end{equation}
Note that, if $\t<0$, the minimal surface ends on the EOW brane at $\t=-\frac{\b}{4}$, and otherwise, it ends at  $\t=\frac{\b}{4}$. These correspond to branches of $m=0,1$ of \eqref{eq:EE_BTZ_w_branch}. The other branches are unphysical. See Fig.\ref{fig:unphysicalBrane}.

Consider the analytical continuation of the entanglement entropy to Lorentzian. First, we focus on the b.c.c. operator with $h_\Psi < c/24$. In this case, $\a_\Psi$ is real and in $(0,1]$. Especially if $\a_\Psi=1$, equivalently $h_\Psi=0$, two boundaries have the same boundary condition, which corresponds to \cite{HartmanMaldacena}. Actually, when $\a_\Psi=1$, the pseudo entropy \eqref{eq:EE_BTZ_w_branch} has no imaginary term and behaves as
\begin{equation}
    S_A=\frac{c}{6}\log\frac{\b}{2\pi\epsilon}+\frac{c}{6}\log\left[2\cosh\frac{2\pi t}{\b}\right]\simeq \frac{c}{6}\frac{2\pi}{\b}t+\frac{c}{6}\log\frac{\b}{2\pi\epsilon}.
\end{equation} Furthermore, even if $\a_\Psi\neq 1$, the real part of the pseudo entropy grows linearly at late times\footnote{We choose the branch of logarithm so that the imaginary part of the pseudo entropy at $t=0$ is zero.}:
\begin{equation}
    S_A^{(\pm)}\simeq \frac{c}{6}\frac{2\pi\a_\Psi}{\b}t + \frac{c}{6}\log\frac{\b}{2\pi\a_\Psi\epsilon} \mp i\frac{c}{12}\pi(\a_\Psi-1).\label{eq:PE_of_BTZ}
\end{equation}
Actually we argue that we need to take the average of the two contributions i.e. 
\be
S_A=\frac{S^{(+)}_A+S^{(-)}_A}{2}, 
\label{entav}
\ee
whose entropies are complex conjugate to each other. This summation is clear from the saddle point approximation in the gravity dual as follows. Since the two contributions 
$S^{(+)}_A$ and $S^{(-)}_A$ are dual to two complex extremal surfaces, whose areas are complex conjugate with each other. Since they have the common real part, they equally contribute to the gravitational path-integral. For the $n$-replicated geometry in AdS$_3$, the partition function $Z^n_G$ in the gravity near $n\simeq 1$ should look like (see also \cite{Kawamoto:2023nki})
\ba
Z^G_n\simeq e^{(1-n)S^{(+)}_A}+e^{(1-n)S^{(-)}_A}.
\ea
The holographic pseudo entropy in the gravity dual is computed as 
\ba
S^{G}_A=\lim_{n\to 1}\frac{1}{1-n}\log \frac{Z^G_n}{(Z^G_1)^n}=\frac{S^{(+)}_A+S^{(-)}_A}{2}+\log 2,
\ea
where we can ignore $\log 2$ in the large $c$ limit.

This cancels the imaginary part and we obtain finally 
\ba
 S_A\simeq \frac{c}{6}\frac{2\pi\a_\Psi}{\b}t + \frac{c}{6}\log\frac{\b}{2\pi\a_\Psi\epsilon}.\label{eq:PE_of_BTZNew}
\ea

Before moving on to the next discussion, we mention that even if we choose any integral path of the geodesic line, the coordinates of the geodesic line must contain complex values, since in our case, the coordinates $X^M$ of the EOW brane are complex.

\subsubsection{Critical phase $h_{\Psi}=\frac{c}{24}$}
From eq. \eqref{eq:EE_Poincare_w_branch}, we find the Lorentzian evolution of the two contributions:
\begin{subequations}
\begin{align}
    S^{(\pm)}_A &= \frac{c}{6}\log\frac{\b}{2\e} + \frac{c}{12}\log\left[1 + \left(\frac{4t}{\b}\right)^2\right]\pm \frac{c}{6}i\tan^{-1}\frac{4t}{\b}\label{eq:poincare_entropy}\\
    &\simeq \frac{c}{6}\log\frac{\b}{2\e} + \frac{c}{6}\log\frac{4t}{\b}\pm\frac{c}{12}\pi i.\label{eq:PE_of_Poin}
\end{align}
\end{subequations}
Again by taking the average (\ref{entav}), which is explained by the gravitational path-integral, we finally obtain the time evolution of pseudo entropy
\be
S_A
\simeq \frac{c}{6}\log\frac{\b}{2\e} + \frac{c}{6}\log\frac{4t}{\b}. \label{eq:PE_of_PoinNEW}
\ee

In this phase, the background geometry is the \Poincare AdS $ds^2=(dz^2-dt^2+dx^2)/z^2$. The EOW branes are placed at $\tau=\pm\b/4$ in the Euclidean geometry. Consider the minimal surface connecting the insertion point $(z,t,x)=(\e,t_0,0)$ of the twist operator to a point on the EOW brane. Let $\t_0:=-it_0$ and suppose $\t_0$ to be real. Then the geodesic line is represented by
\begin{equation}
    z(s)=(\b/4-\t_0)\sin s,\quad \t(s)=\b/4-(\b/4-\t_0)\cos s,\quad s:s_\e\mapsto \pi/2,
\end{equation}
where $s_\e$ is defined via cutoff $\e$ as
\begin{equation}\label{eq:s_epsilon}
    \e =z(s_\e)=(\b/4-\t_0)\sin s_\e.
\end{equation}

This geodesic length is
\begin{equation}
    L=\int_{s_\e}^{\pi/2}\frac{\sqrt{z^\prime(s)^2+\tau^\prime(s)^2}}{z(s)}ds=\int_{s_\e}^{\pi/2}\frac{ds}{\sin s}=-\log\tan\frac{s_\e}{2}
\end{equation}

Note that since the integrand function has poles $s=n\pi,(n\in\ZZ)$, the geodesic length is independent of the integral path unless the path crosses any of these poles. Using eq.\eqref{eq:s_epsilon}, the geodesic length $L$ becomes
\begin{equation}
    L\simeq\log\frac{\b-4it_0}{2\e}
\end{equation}
This result coincides with the entropy eq.\eqref{eq:poincare_entropy}.

Using this parametrisation, we can check more explicitly that the coordinates of the Ryu-Takayanagi surface have complex values.

if $z(s) = (\b/4 + it_0)$ is real, $\sin(s)$ is represented as $\sin(s) = r(s)(\b/4 -it_0)$, where $r(s)$ is real. Hence $t(s)$ is
\begin{equation}
    -it(s) = \b/4 - \sqrt{(\b/4+it_0)^2-r^2(s)(\b^2/4^2+t^2_0)},
\end{equation}
Therefore
\begin{equation}
    (it(s)+\b/4)^2 = -i(t_0-t(s))(\b/2-i(t_0+t(s))).
\end{equation}
If $t(s)$ is real, the equation above does not hold. Since, the coordinates is complex

\subsubsection{Large deformation phase $h_{\Psi}>\frac{c}{24}$}
In this phase, the bulk dual becomes the thermal AdS
\begin{equation}
    ds^2 = \frac{dz^2}{z^2h(z)} + \frac{d\tau^2}{z^2} + \frac{h(z)dx^2}{z^2},\quad h(z) = 1-\frac{z^2}{a^2}.
\end{equation}
Usually, the thermal AdS space is compactified in $x$-direction $x\sim x + 2\pi a$ in order to remove a conical singularity at $z=a$. However, since we are focusing on the infinite strip, which is non-compact, the holographic dual geometry is interpreted as an universal covering of thermal AdS. 
This implies that the holographic dual has a naked singularity at $z=a$, but it is not a conical singularity in the usual sense, since one cannot go around $z=a$ because of the non-compactness. This type of singularity is not strange and appears when heavy operators above the black hole threshold are inserted \cite{Abajian:2023bqv,Abajian:2023bqv,Tian:2024fmo}. This is exactly our case: while the geometry we are looking at is thermal AdS from the closed string point of view, it is a BTZ black hole from the open string point of view.  

\begin{figure}[htbp]
    \centering
    \includegraphics[width=0.6\linewidth]{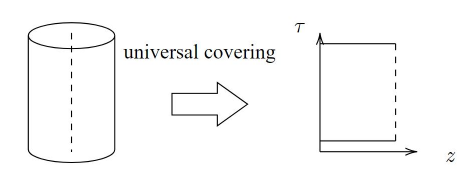}
    \vspace{1cm}
      \includegraphics[width=0.2\linewidth]{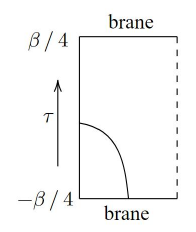}
    \caption{A sketch of gravity dual of the large deformation phase (left) and the profile of geodesic (right). This dual geometry is an universal covering of usual thermal AdS.}
    \label{fig:enter-label}
\end{figure}

In usual thermal AdS, the tension-less EOW brane is simply $\t$-constant plane. Even if taking universal covering, this property is inherited.

Therefore, the holographic entanglement entropy is obtained by find the geodesic distance between a point $(z,\t,x)=(\e,\t,0)$ and the EOW brane $\t = \pm \b/4$:
\begin{equation}
    S_A = \frac{c}{6}\cosh^{-1}\left[\frac{a}{\e}\sinh\left(\frac{\b/4 - \abs{\t}}{a}\right)\right] \simeq \frac{c}{6}\log\frac{a}{\epsilon } +\frac{c}{6}\log\left[ 2\sinh\left(\frac{\beta /4-| \tau | }{a}\right)\right].
\end{equation}
By identifying $a = \frac{\b}{2\pi\a_\Psi}$, this holographic entanglement entropy coincides with the entanglement entropy in large deformation phase.

Similarly, we obtain the Lorentzian time evolution via the Wick rotation:
\begin{align}
 S^{(\pm)}_A &= \frac{c}{6}\log\frac{\b}{2\pi\abs{\a_\Psi}\e} + \frac{c}{12}\log\left[2\left(\cosh\pi\abs{\a_\Psi}-\cos\frac{4\pi\abs{\a_\Psi}t}{\b}\right)\right]\nonumber\\
    &\quad\quad\mp\frac{c}{12}i\tan^{-1}\left[\frac{\sin(4\pi\abs{\a_\Psi}t/\b)\sinh\pi\abs{\a_\Psi}}{\cos(4\pi\abs{\a_\Psi}t/\b)\cosh\pi\abs{\a_\Psi}-1}\right].\label{eq:PE_of_TAdS_full}
\end{align}
Again by taking the average (\ref{entav}), we finally obtain 
\begin{align}
 S_A &= \frac{c}{6}\log\frac{\b}{2\pi\abs{\a_\Psi}\e} + \frac{c}{12}\log\left[2\left(\cosh\pi\abs{\a_\Psi}-\cos\frac{4\pi\abs{\a_\Psi}t}{\b}\right)\right].
   \label{eq:PE_of_TAdS_fullNEW}
\end{align}
Especially, if $\abs{\a_\Psi}\gg 1$, eq.\eqref{eq:PE_of_TAdS_fullNEW} can be approximated as the constant value:
\begin{equation}
    S_A \simeq \frac{c}{6}\log\frac{\b}{2\pi\abs{\a_\Psi}\e} + \frac{c}{12}\pi\abs{\a_\Psi}.\label{eq:PE_of_TAdS_large_a}
\end{equation}
In this large deformation, we again find that the coordinate of the geodesic which computes the holographic pseudo entropy gets complex valued.

\section{Pseudo entropy in Dirac fermions CFT}\label{sec:free}

So far, as the main result of this paper, we have seen that the pseudo entropy in holographic CFTs, shows an entanglement phase transition  behavior as we increase the difference between two boundary conditions. Here, before we conclude this paper, we would like to examine what will happen for a CFT with the opposite property, by taking the free Dirac fermion CFT as such an example. As we will see below, in this case the result for the different boundary conditions gets identical to that for the same boundary condition, as opposed to the holographic CFT result.

We consider the $c=1$ free massless Dirac fermion CFT on a two dimensional cylinder whose coordinate is written as $w=\tau+i\sigma$ and $\bar{w}=\tau-i\sigma$, where $\tau$ is the Euclidean time and the space coordinate $\sigma$ is compactified as $\sigma\sim \sigma+2\pi$. We describe the Dirac fields as $(\Psi_L(w),\bar{\Psi}_L(w))$ and $(\Psi_R(\bar{w}),\bar{\Psi}_R(\bar{w}))$ in the left and right-moving modes, respectively. We would like to compute the pseudo entropy where the initial state and final state are given by two different boundary states $|N\lb$ and $|D\lb$, called the Neumann and Dirichlet boundary condition. They are defined by the following conditions at $\tau=0$ for any $\sigma$:
\ba
&&  \left(\Psi_L(w)-\Psi_R(\bar{w})\right)|N\lb=0,\no
&&  \left(\Psi_L(w)-\bar{\Psi}_R(\bar{w})\right)|D\lb=0.\label{bcgfh}
\ea
When the initial and final state are given by the same boundary state (Dirichlet or Neumann b.c.), the entanglement entropy was computed in \cite{Takayanagi:2010wp} via the replica method \cite{Casini:2005rm,Azeyanagi:2007bj}.  We can apply a similar analysis to our pseudo entropy calculation by choosing the one of the two boundary conditions is Dirichlet and the other is Neumann. We presented the details in the appendix A and below we simply write only the outline and the final result. 

The basic strategy, as summarized in Appendix A, is as follows. First we consider $n$ sheets of the replica, in order to calculate $\mbox{Tr}[(\rho_A)^n]$. We have $n$ copies of the fermion
\ba
(\Psi^{(k)}_L(w),\bar{\Psi}^{(k)}_L(w)),\ \ 
(\Psi^{(k)}_R(\bar{w}),\bar{\Psi}^{(k)}_R(\bar{w})), 
\ea
where
$k=1,2,\ddd,n$. Then we perform a discrete Fourier transformation into to the other fermion basis 
\ba
(\psi^{(a)}_L(w),\bar{\psi}^{(a)}_L(w)),\ \  (\psi^{(a)}_R(\bar{w}),\bar{\psi}^{(a)}_R(\bar{w})), 
\label{dfermi}
\ea
with $a=-\frac{n-1}{2},\ddd,\frac{n-1}{2}$, to make the boundary condition at the end points of $A$ diagonal. Then we bosonize the fermions (\ref{dfermi}) into free scalar fields 
\ba
\phi^{(a)}_L(w),\  \phi^{(a)}_R(\bar{w})
\label{dscalar}
\ea
Then we can construct the twist operator for the replica method by 
\ba
&& \sigma_n(w_1,\bar{w}_1)=\prod^{\frac{n-1}{2}}_{a=-\frac{n-1}{2}}e^{i\frac{a}{N}(\phi^{(a)}_L(w_1)-\phi^{(a)}_R(\bar{w}_1))},\no
&& \bar{\sigma}_n(w_2,\bar{w}_2)=\prod^{\frac{n-1}{2}}_{a=-\frac{n-1}{2}}e^{-i\frac{a}{N}(\phi^{(a)}_L(w_2)-\phi^{(a)}_R(\bar{w}_2))}.\label{twitssc}
\ea
Even though in \cite{Takayanagi:2010wp}, the above one for the Neumann boundary condition and another different one for the Dirichlet boundary condition, were employed for the replica calculation, we can simply use the above one for both Neumann and Dirichlet case and obtain the correct result, as we show in (\ref{showtt}) of appendix A.

By choosing the subsystem $A$ to be the interval $[w_1,w_2]$, we can write the partition function on the replicated space as follows
\ba
\mbox{Tr}[(\rho_A)^n]=\frac{\la B_1|e^{-2\ep H}\sigma_n(w_1)\bar{\sigma}_n(w_2)|B_2\lb}{\la B_1|e^{-2\ep H}|B_2\lb},
\ea
where $\ep$ is the regularization of the boundary state and $H$ is the CFT Hamiltonian. In \cite{Takayanagi:2010wp}, the cases $|B_1\lb=|B_2\lb=|N\lb$ and $|B_1\lb=|B_2\lb=|D\lb$ were explicitly computed by evaluating the above correlation function. 

We can apply a similar analysis to our pseudo entropy calculation by choosing $|B_1\lb=|D\lb$ and $|B_2\lb=|N\lb$. We presented the details in the appendix A. In the end, we obtain the final expression, given by 
\ba
&&\mbox{Tr}(\rho_A)^n \no
&&=\prod_{a=1}^{\frac{n-1}{2}}\left[\frac{\theta_1\left(\frac{\ep +it}{\pi}+\frac{\sigma}{2\pi}|\frac{2i\ep}{\pi}\right)\theta_1\left(\frac{\ep +it}{\pi}-\frac{\sigma}{2\pi}|\frac{2i\ep}{\pi}\right)\eta\left(\frac{2i\ep}{\pi}\right)^6}{\theta_1\left(\frac{\sigma}{2\pi}|\frac{2i\ep}{\pi}\right)^2\theta_1\left(\frac{\ep +it}{\pi}|\frac{2i\ep}{\pi}\right)^2}\right]^{\frac{2a^2}{n^2}}.  \label{resultndad}
\ea
This is identical to the result of \cite{Takayanagi:2010wp}, where the boundary conditions are both Neumann or Dirichlet. Thus we can conclude that the pseudo entropy in the free Dirac fermion CFT is identical to the entanglement entropy for the Dirichlet or Neumann boundary condition. In this way we do not observe any non-trivial behavior of the pseudo entropy in this free CFT as opposed to the result in holographic CFTs. We expect that this is owing to the integrable structure of the free CFT.

\section{Conclusions and Discussions}

In this paper, we analyzed the time evolution of pseudo entropy for two quantum states constructed by the regularized boundary states in BCFTs. The pseudo entropy is an extension of entanglement entropy to a setup with a post-selection process. The pseudo entropy depends on two different quantum states, namely the initial and final state. In a two dimensional holographic CFT, we considered two different boundary states $|B_a\lb$ and $|B_b\lb$, which are intertwined by boundary condition changing operators $\Psi_{ab}$ and $\Psi_{ba}$. 
The pseudo entropy, in the replica method, is evaluated by applying the mirror method in BCFTs and the heavy-heavy-light-light approximation of the conformal block in large $c$ CFTs.

This analysis leads to the result that when the conformal dimension $h_\Psi$ of the boundary condition changing operator is small $h_{\Psi}< \frac{c}{24}$, the pseudo entropy grows linearly $S_A\sim \frac{c}{6}\cdot\frac{2\pi\ap_\Psi}{\beta}\cdot t$ at late time. However when it is larger  $h_{\Psi}> \frac{c}{24}$, the pseudo entropy stays constant. At the critical value $h_{\Psi}= \frac{c}{24}$, the pseudo entropy grows logarithmically $S_A\simeq \frac{c}{6}\log \frac{t}{\beta}$. This result for a holographic CFT, qualitatively agrees with those obtained in our earlier work \cite{Kanda:2023zse,Kanda:2023jyi}, where we studied the same problem by employing the AdS/BCFT model, which provides a bottom up effective model for a gravity dual of a holographic BCFT, assuming that the two states $|B_\ap\lb$ and $|B_\beta\lb$ are related by an exactly marginal perturbation. An interesting difference between the present one and the previous one \cite{Kanda:2023zse,Kanda:2023jyi} is that the coefficient of $\log t$ growth at the critical point is given by $\frac{c}{6}$ in the former and by $\frac{c}{3}$ in the latter. 
This difference is not contradictory but rather expected. In \cite{Kanda:2023zse,Kanda:2023jyi}, the changing of the boundary condition is mediated by a marginal deformation localized on the boundary, where it was assumed that the dual scalar field is confined on the EOW brane in AdS. On the other hand, the boundary conditions are changed by a heavy boundary primary operator in the spectrum of the open string, which is dual to a conical defect or spacetime banana \cite{Abajian:2023jye,Abajian:2023bqv,Tian:2024fmo}, depending on its conformal weight. In short, they have different descriptions in AdS gravity models. A quick way to see the difference is through the one-point function of a heavy bulk scalar primary with $h=\bar{h} < c/24$. If we consider the gravity model where a heavy bulk particle cannot end of the EOW brane as discussed in \cite{Kawamoto:2022etl,Kusuki:2022ozk,Wang:2025bcx}, then the changing of the boundary condition in the way discussed in \cite{Kanda:2023zse,Kanda:2023jyi} will give a vanishing one-point function for a heavy bulk scalar primary. On the other hand, the changing of the boundary condition in the way discussed in the current paper will give a non-vanishing one-point function, due to the interaction between the particle induced by the bulk primary and that induced by the boundary condition changing primaries.

Moreover, it is also intriguing to note that the above behavior is qualitatively the same as those observed in the entanglement phase transition (or measurement induced phase transition) \cite{Li:2018mcv,Skinner:2018tjl,Li:2019zju,Kawabata:2022biv}. It would be intriguing to make this connection clearer in future works.

In addition, we also computed the pseudo entropy in a Dirac free fermion CFT when the two different quantum states are given by the Dirichlet and Neumann boundary state, respectively. In this case we found that the result is identical to the entanglement entropy for the Dirichlet or Neumann boundary state. Therefore the effect of the boundary changing is invisible here. This may make us wonder if this non-trivial phase transition behavior we found for the holographic CFTs, may be tied with its irrational or chaotic property. It is an interesting future direction to explore more on what kind of CFTs can exhibit such a phase transition of pseudo entropy.

\section*{Acknowledgements}
We are grateful to Yuya Kusuki and Shinsei Ryu for useful discussions. This work is supported by by MEXT KAKENHI Grant-in-Aid for Transformative Research Areas (A) through the ``Extreme Universe'' collaboration: Grant Number 21H05187. TT is also supported by Inamori Research Institute for Science and by JSPS Grant-in-Aid for Scientific Research (B) No.~25K01000.

\appendix
\section{Computations in Dirac Fermion CFT}

Here we present the detailed calculations of pseudo entropy in the free massless Dirac fermion CFT in the presence of Dirichlet and Neumann boundary condition (\ref{bcgfh}), whose setup and outline was presented in section \ref{sec:free}.  We will follow the convention of this paper \cite{Takayanagi:2010wp} below.

\subsection{Replica method and bosonization}

We consider the setup where the initial state is given by $e^{-\ep H}|N\lb$ at $t=0$ and the final state by $e^{-\ep H}|D\lb$ at $t=0$ and measure the pseudo entropy at the time $t=t_0$ for the subsystem $A$ given by an interval $[w_1,w_2]$. This setup is given by a cylinder length $2\ep$ with the twist operator $\sigma_n$ and $\bar{\sigma}_n$ inserted at the two end points of the subsystem $A$:
\ba
(w_1,\bar{w}_1)=(\ep+it_0+i\sigma_1,\ep+it_0-i\sigma_1),
\ea
and 
\ba
(w_2,\bar{w}_2)=(\ep+it_0+i\sigma_2,\ep+it_0-i\sigma_2),
\ea
respectively.  Thus we want to calculate 
\ba
\mbox{Tr}(\rho_A)^n=\la D_n|e^{-2\ep H}\sigma_n(w_1,\bar{w}_1)\bar{\sigma}_n(w_2,\bar{w}_2)|N_n\lb,\label{trn}
\ea
where $|N_n\lb$ and $|D_n\lb$ are the boundaries states for the $n$ replicated CFT, constructed as follows. Also note $H=L_0+\ti{L}_0-\frac{n}{12}$ is the Hamiltonian of the CFT.

We write $n$ replicated Dirac fermion fields as $\Psi^{(k)}_{L,R}$ and $\bar{\Psi}^{(k)}_{L,R}$ for $k=0,1,2,\ddd,n-1$. We assume $n$ is an odd integer. Then the boundary conditions are given by
\ba
&&  \left(\Psi^{(k)}_L(w)-\Psi^{(k)}_R(\bar{w})\right)|N_n\lb=0,\no
&&  \left(\Psi^{(k)}_L(w)-\bar{\Psi}^{(k)}_R(\bar{w})\right)|D_n\lb=0.
\label{bscond}
\ea
For the replica method, we need the following boundary conditions at $w=w_1$ and 
$w=w_2$, i.e. the end points of the subsystem $A$:
\ba
&& \Psi^{(k)}_L(we^{2\pi i})=\Psi^{(k+1)}_L(w),\ \ \ \  \bar{\Psi}^{(k)}_L(we^{2\pi i})=\bar{\Psi}^{(k+1)}_L(w),\no
&& \Psi^{(k)}_R(\bar{w}e^{-2\pi i})=\Psi^{(k+1)}_R(\bar{w}),\ \ \ \  \bar{\Psi}^{(k)}_R(\bar{w}e^{-2\pi i})=\bar{\Psi}^{(k+1)}_R(\bar{w}).
\label{twistcbc}
\ea

Now we introduce the discrete Fourier transformation as follows:
\ba
&& \psi^{(a)}_L=\sum_{k=0}^{n-1}e^{-2\pi i\frac{ak}{n}}\Psi^{(k)}_L,\ \ \ \ \bar{\psi}^{(a)}_L=\sum_{k=0}^{n-1}e^{2\pi i\frac{ak}{n}}\bar{\Psi}^{(k)}_L,\no
&& \psi^{(a)}_R=\sum_{k=0}^{n-1}e^{-2\pi i\frac{ak}{n}}\Psi^{(k)}_R,\ \ \ \ \bar{\psi}^{(a)}_R=\sum_{k=0}^{n-1}e^{2\pi i\frac{ak}{n}}\bar{\Psi}^{(k)}_R,
\ea
where $a$ takes $n$ different values: $a=-\frac{n-1}{2},-\frac{n-3}{2},\ddd,0,\ddd,\frac{n-3}{2},\frac{n-1}{2}$. After this transformation, the twisted boundary condition (\ref{twistcbc}) is rewritten as 
\ba
&& \psi^{(a)}_L(we^{2\pi i})=e^{2\pi i\frac{a}{N}}\psi^{(a)}_L(w),\ \ \ \
\bar{\psi}^{(a)}_L(we^{2\pi i})=e^{-2\pi i\frac{a}{N}}\bar{\psi}^{(a)}_L(w),\no
&& \psi^{(a)}_R(\bar{w}e^{-2\pi i})=e^{2\pi i\frac{a}{N}}\psi^{(a)}_R(\bar{w}),\ \ \ \ \bar{\psi}^{(a)}_R(\bar{w}e^{-2\pi i})=e^{-2\pi i\frac{a}{N}}\bar{\psi}^{(a)}_R(\bar{w}). \label{twistbg}
\ea
We now bosonize the Dirac fermion as follows
\ba
\psi^{(a)}_L(w)=e^{i\phi^{(a)}_L(w)},\ \ \ \ \bar{\psi}^{(a)}_L(w)=e^{-i\phi^{(a)}_L(w)}, \no
\psi^{(a)}_R(\bar{w})=e^{i\phi^{(a)}_R(\bar{w})},\ \ \ \ \bar{\psi}^{(a)}_R(\bar{w})=e^{-i\phi^{(a)}_R(\bar{w})}.
\ea

Then we can realize the expected twisted boundary condition (\ref{twistbg}) can be realized by choosing the twist operators as 
in (\ref{twitssc}). 

Now the boundary condition of the boundary states (\ref{bscond}) can be expressed as 
\ba
&&  \left(\psi^{(a)}_L(w)-\psi^{(a)}_R(\bar{w})\right)|N_n\lb=0,\no
&&  \left(\psi^{(a)}_L(w)-\bar{\psi}^{(-a)}_R(\bar{w})\right)|D_n\lb=0.
\label{bscondf}
\ea
Notice that in the Dirichlet case, the above boundary condition mixes two different fermions. 

\subsection{Scalar field description}

The massless free scalar fields $\phi^{(a)}_{L,R}$ are expanded as follows
\ba
&& \phi^{(a)}_L(w)=x^{(a)}_L-ip^{(a)}_Lw+i\sum_{m\neq 0}\frac{\ap^{(a)}_m}{m}
e^{-mw},\no
&& \phi^{(a)}_R(\bar{w})=x^{(a)}_R-ip^{(a)}_R w+i\sum_{m\neq 0}\frac{\ti{\ap}^{(a)}_m}{m}e^{-m\bar{w}},
\ea
The zero modes are quantized as follows
\ba
p^{(a)}_L=\frac{n^a}{R}+\frac{w^aR}{2},\ \ \ \ p^{(a)}_R=\frac{n^a}{R}-\frac{w^aR}{2}.
\ea
In order to be equivalent to Dirac fermion CFT we set $R=1$.

The boundary state condition (\ref{bscondf}) is now rewritten in terms of free scalars as follows (for any $a$):
\ba
&& \left(\phi^{(a)}_L(w)-\phi^{(a)}_R(\bar{w})\right)|N_n\lb=0,\no
&& \left(\phi^{(a)}_L(w)+\phi^{(-a)}_R(\bar{w})\right)|D_n\lb=0. \label{bosonbc}
\ea
In terms of the oscillators, we find from (\ref{bosonbc})
\ba
&& (\ap^{(a)}_{m}+\ti{\ap}^{(-a)}_{-m})|N_n\lb=0,\no
&& (\ap^{(a)}_{m}-\ti{\ap}^{(-a)}_{-m})|D_n\lb=0.
\ea
For the zero modes we find the constraints
\ba
&& n^{(a)}|N_n\lb=0,\no
&& (n^{a}-n^{(-a)})|D_n\lb=(w^{a}+w^{(-a)})|D_n\lb=0.
\ea

These were solved as follows:
\ba
&& |N_n\lb={\cal N}_{1}\prod^{\frac{n-1}{2}}_{a=-\frac{n-1}{2}}\left(\exp\left[-\sum^\infty_{n=1}\frac{1}{n}(\ap^{(a)}_{-n}\ti{\ap}^{(a)}_{-n})\right]\sum_{w^a\in Z}|w^a\lb\right),\no
&& |D_n\lb={\cal N}_{2}\prod^{\frac{n-1}{2}}_{a=1}\left(\exp\left[\sum^\infty_{n=1}\frac{1}{n}(\ap^{(a)}_{-n}\ti{\ap}^{(-a)}_{-n}+\ap^{(-a)}_{-n}\ti{\ap}^{(a)}_{-n})\right]\sum_{w^a, n^a\in Z}|n^a,w^a\lb|-n^a,-w^a\lb\right)
\otimes |D_0\lb,\label{bssc}
\ea
where we defined $|D_0\lb$ as follows:
\ba
|D_0\lb=\exp\left[-\sum^\infty_{n=1}\frac{1}{n}(\ap^{(0)}_{-n}\ti{\ap}^{(0)}_{-n})\right]\sum_{n^0\in Z}|n^0\lb.
\ea

For the Neumann b.c. , the zero mode summation reads 
\ba
&& \sum_{w=-\infty}^\infty \la w| e^{i\frac{a}{N}\left[\phi^{(a)}_L(w_1)-\phi^{(a)}_R(\bar{w_1})\right]}e^{-i\frac{a}{N}\left[\phi^{(a)}_L(w_2)-\phi^{(a)}_R(\bar{w_2})\right]}|w\lb.\no
&& =\sum_w e^{-\frac{R^2}{2}w^2\ep} e^{\frac{a^2}{2N^2}(w_2-w_1+\bar{w}_2-\bar{w}_1)}.
\label{wsumd}
\ea
Note that there is no $\sigma$ dependence in the summation over $w$, which makes the zero model contribution to the EE trivial. 

For the Dirichlet b.c. ,  the zero model summation over the momentum reads 
\ba
&& \sum_{n=-\infty}^\infty \la n| e^{i\frac{a}{N}\left[\phi^{(a)}_L(w_1)-\phi^{(a)}_R(\bar{w_1})\right]}e^{-i\frac{a}{N}\left[\phi^{(a)}_L(w_2)-\phi^{(a)}_R(\bar{w_2})\right]}|n\lb.\no
&& =\sum_w e^{-\frac{2}{R^2}n^2\ep} e^{\frac{na}{NR}(w_2-w_1-\bar{w}_2+\bar{w}_1)}e^{\frac{a^2}{2N^2}(w_2-w_1+\bar{w}_2-\bar{w}_1)}.
\label{showtt}
\ea
However, actually we need to consider the summation over $n^a$ and $w^a$ for the state $|n^a,w^a\lb_a|-n^a,-w^a\lb_{-a}$, the $n^a$ contribution and $-n^a$ contribution cancells for the factor $e^{\frac{na}{NR}(w_2-w_1-\bar{w}_2+\bar{w}_1)}$. Moreover the summation over 
$w^a$ does not produce such a factor which depends on the coordinate as in (\ref{wsumd}). Thus again the zero model contributions is trivial in the Dirichlet case. In this way, we can understand that the EE for the Nuemann b.c. is equal to that for the Dirichlet one, by using the same twist operator (\ref{twitssc}) (i.e. $\sigma^{(1)}$ in \cite{Takayanagi:2010wp}).

\subsection{Computing pseudo entropy}
Now let us evaluate (\ref{trn}) by employing the free scalar representation 
(\ref{twitssc}) and (\ref{bssc}). We focus on the sector of $a$ and $-a$ for a fixed value of $a$ as they are coupled and we will take the sum over $a$ only in the final stage. The $a=0$ sector is special 

First, the zero mode part is computed as follows (there is a typo in (A.41) of \cite{Takayanagi:2010wp}):
\ba
&& \sum_{w=-\infty}^\infty \la w|\la -w| e^{i\frac{a}{N}\left[\phi^{(a)}_L(w_1)-\phi^{(a)}_R(\bar{w_1})-\phi^{(-a)}_L(w_1)+\phi^{(-a)}_R(\bar{w_1})\right]}e^{-i\frac{a}{N}\left[\phi^{(a)}_L(w_2)-\phi^{(a)}_R(\bar{w_2})-\phi^{(-a)}_L(w_2)+\phi^{(-a)}_R(\bar{w_1})\right]}|w\lb |-w\lb.\no
&& =\sum_w e^{-R^2w^2\ep} e^{\frac{a^2}{n^2}(w_2-w_1+\bar{w}_2-\bar{w}_1)}.
\ea

Next we turn to the oscillator modes. We write 
\ba
&& \frac{\ap^{(a)}_m}{\s{m}}\to \ap,\ \ \ \ \frac{\ap^{(a)}_{-m}}{\s{m}}\to \ap^\dagger, \no
&& \frac{\ti{\ap}^{(a)}_m}{\s{m}}\to \ti{\ap},\ \ \ \ \frac{\ti{\ap}^{(a)}_{-m}}{\s{m}}\to \ti{\ap}^\dagger, \no
&& \frac{\ap^{(-a)}_m}{\s{m}}\to \beta,\ \ \ \ \frac{\ap^{(-a)}_{-m}}{\s{m}}\to \beta^\dagger,\no
&& \frac{\ti{\ap}^{(-a)}_m}{\s{m}}\to \ti{\beta},\ \ \ \ \frac{\ti{\ap}^{(-a)}_{-m}}{\s{m}}\to \ti{\beta}^\dagger.
\ea
We will postpone taking the product over $m$ in the end.

The inner product  (\ref{trn}) for the fixed mode $m$, denoted by $I_m$, looks like (we set $\lambda\equiv \frac{a}{n\s{m}}$)
\ba
&& I_m=\la 0|e^{(\ap\ti{\beta}+\ti{\ap}\beta)e^{-4\ep m}}e^{-\lambda\left[(\ap-\beta)e^{-mw_1}-(\ap^\dagger-\beta^\dagger)e^{mw_1}-(\ti{\ap}-\ti{\beta})e^{-m\bar{w}_1}+(\ti{\ap}^\dagger-\ti{\beta}^\dagger)e^{m\bar{w}_1}\right]}\no
&&\ \ \ \ \times  e^{\lambda\left[(\ap-\beta)e^{-mw_2}-(\ap^\dagger-\beta^\dagger)e^{mw_2}-(\ti{\ap}-\ti{\beta})e^{-m\bar{w}_2}+(\ti{\ap}^\dagger-\ti{\beta}^\dagger)e^{m\bar{w}_2}\right]}e^{-(\ap^\dagger\ti{\ap}^\dagger+\beta^\dagger\ti{\beta}^\dagger)}|0\lb\no
&&=\la 0|e^{(a\ti{a}-b\ti{b})e^{-4\ep m}}e^{\s{2}\lambda b^\dagger e^{mw_1}}e^{-\s{2}\lambda \ti{b}^\dagger e^{m\bar{w}_1}}e^{-\s{2}\lambda b e^{-mw_1}}e^{\s{2}\lambda \ti{b} e^{-m\ti{w}_1}}\no
&&\ \ \ \ \times e^{-\s{2}\lambda b^\dagger e^{mw_2}}e^{\s{2}\lambda \ti{b}^\dagger e^{m\bar{w}_2}}e^{\s{2}\lambda b e^{-mw_2}}e^{-\s{2}\lambda \ti{b} e^{-m\ti{w}_2}}
e^{-(a^\dagger\ti{a}^\dagger+b^\dagger\ti{b}^\dagger)}|0\lb\no
&& =\frac{1}{1+e^{-4m\ep}}\cdot 
e^{8\lambda^2-2\lambda^2(e^{m(w_1-w_2)}+e^{m(\bar{w}_1-\bar{w}_2)})}\no
&&\times\ \la 0|e^{-b\ti{b}e^{-4\ep m}}e^{-\s{2}\lambda(e^{-mw_1}-e^{-mw_2})b
-\s{2}\lambda(e^{-m\bar{w}_2}-e^{-m\bar{w}_1)\ti{b}}}e^{\s{2}\lambda(e^{mw_1}-e^{mw_2})b^\dagger
+\s{2}\lambda(e^{m\bar{w}_2}-e^{m\bar{w}_1)\ti{b}^\dagger}}e^{-b^\dagger \ti{b}^\dagger}|0\lb.\no
\ea
The total contribution from $a$ and $-a$ sector to (\ref{trn}) is written as (setting $R=1$)
\ba
\left[\mbox{Tr}(\rho_A)^n\right]_{a,-a}=\sum_w e^{-w^2\ep} e^{\frac{a^2}{n^2}(w_2-w_1+\bar{w}_2-\bar{w}_1)}\cdot e^{\frac{\ep}{3}}\prod_{m=1}^\infty I_m,\label{tota}
\ea
where the factor $e^{\frac{\ep}{3}}$ comes from the Casimir energy term.

We can evaluate the above inner product by using (A.42) of \cite{Takayanagi:2010wp} or (D.11) of \cite{Kawamoto:2025oko}. This leads to 
\ba
I_m=\frac{1}{1-e^{-8m\ep}}\cdot e^{8\lambda^2-2\lambda^2(e^{m(w_1-w_2)}+e^{m(\bar{w}_1-\bar{w}_2)})}\cdot e^{K_m},
\ea
where $K_m$ is computed as 
\ba
&& K_m\no
&& =\frac{2a^2}{n^2m(1-e^{-4m\ep})}\Bigl(-4+e^{m(w_1-w_2)}+e^{m(w_2-w_1)}+
e^{m(\bar{w}_1-\bar{w}_2)}+e^{m(\bar{w}_2-\bar{w}_1)}+e^{-m(w_1+\bar{w}_1)}+e^{-m(w_2+\bar{w}_2)}\no
&& \ \ -e^{-m(w_1+\bar{w}_2)}+e^{-m(w_2+\bar{w}_1)}+e^{-4m\ep}\left(e^{m(w_1+\bar{w}_1)}+e^{m(w_2+\bar{w}_2)}-e^{m(w_1+\bar{w}_2)}+e^{m(w_2+\bar{w}_1)}\right)\Bigr).
\ea
By expanding $\frac{1}{1-e^{-4m\ep}}$ as $\sum_{p=1}^\infty e^{-4pm\ep}$ and using the formula $\sum_{m=1}^{\infty}\frac{y^m}{m}=-\log (1-y)$, we can evaluate the product over $m=1,2,\ddd$ as follows:
\ba
\log \prod_{m=1}^\infty I_m&=& -\sum_{p=1}^\infty\log (1-e^{-8p\ep})+\frac{2a^2}{n^2}\log(1-e^{w_1-w_2})+\frac{2a^2}{n^2}\log(1-e^{\bar{w}_1-\bar{w}_2})\no
&&+\frac{8a^2}{n^2}\sum_{p=1}^\infty\log(1-e^{-4p\ep})+\frac{2a^2}{n^2}\sum^\infty_{p=0}J_p,
\ea
where $J_p$ is defined by
\ba
&& J_p=-\log\left[(1-e^{-4p\ep+(w_1-w_2)})(1-e^{-4p\ep-(w_1-w_2)})\right]
-\log\left[(1-e^{-4p\ep+(\bar{w}_1-\bar{w}_2)})(1-e^{-4p\ep-(\bar{w}_1-\bar{w}_2)})\right]\no
&&\ \ \ \ \ \ -\log\left[(1-e^{-4p\ep+(w_1+\bar{w}_1)})(1-e^{-4p\ep-(w_1+\bar{w}_1)})\right]
-\log\left[(1-e^{-4p\ep+(w_2+\bar{w}_2)})(1-e^{-4p\ep-(w_2+\bar{w}_2)})\right]\no
&&\ \ \ \ \ \ +\log\left[(1-e^{-4p\ep+(w_1+\bar{w}_2)})(1-e^{-4p\ep-(w_1+\bar{w}_2)})\right]
+\log\left[(1-e^{-4p\ep+(w_2+\bar{w}_1)})(1-e^{-4p\ep-(w_2+\bar{w}_1)})\right].\no
\ea

Finally the total contribution (\ref{tota}) in the $a$ and $-a$ sector is found to be 
\ba
&& \left[\mbox{Tr}(\rho_A)^n\right]_{a,-a}\no
&&={\cal N}_a \cdot\left[\frac{\sum_{w=-\infty}^\infty e^{-w^2\ep}}{\eta\left(\frac{4i\ep}{\pi}\right)}\right]\cdot 
\left[\frac{\theta_1\left(\frac{\ep +it}{\pi}+\frac{\sigma}{2\pi}|\frac{2i\ep}{\pi}\right)\theta_1\left(\frac{\ep +it}{\pi}-\frac{\sigma}{2\pi}|\frac{2i\ep}{\pi}\right)\eta\left(\frac{2i\ep}{\pi}\right)^6}{\theta_1\left(\frac{\sigma}{2\pi}|\frac{2i\ep}{\pi}\right)^2\theta_1\left(\frac{\ep +it}{\pi}|\frac{2i\ep}{\pi}\right)^2}\right]^{\frac{2a^2}{n^2}},
\ea
where we introduced $\sigma\equiv\sigma_2-\sigma_1$. The coefficient ${\cal N}_a$ is the normalization factor and we would like to normalize such that the total expression is divided by the vacuum partition function on the cylinder. Thus we set 
\ba
{\cal N}_a \cdot\left[\frac{\sum_{w=-\infty}^\infty e^{-w^2\ep}}{\eta\left(\frac{4i\ep}{\pi}\right)}\right]=1.
\ea
Thus we obtain 
\ba
&& \left[\mbox{Tr}(\rho_A)^n\right]_{a,-a}=
\left[\frac{\theta_1\left(\frac{\ep +it}{\pi}+\frac{\sigma}{2\pi}|\frac{2i\ep}{\pi}\right)\theta_1\left(\frac{\ep +it}{\pi}-\frac{\sigma}{2\pi}|\frac{2i\ep}{\pi}\right)\eta\left(\frac{2i\ep}{\pi}\right)^6}{\theta_1\left(\frac{\sigma}{2\pi}|\frac{2i\ep}{\pi}\right)^2\theta_1\left(\frac{\ep +it}{\pi}|\frac{2i\ep}{\pi}\right)^2}\right]^{\frac{2a^2}{n^2}},
\ea

Finally the total contribution (\ref{trn}) to Tr$(\rho_A)^n$ reads 
\ba
&&\mbox{Tr}(\rho_A)^n \no
&&=\left(\prod_{a=1}^{\frac{n-1}{2}}\left[\mbox{Tr}(\rho_A)^n\right]_{a,-a}\right)\cdot \left[\mbox{Tr}(\rho_A)^n\right]_{0}\no
&&=\prod_{a=1}^{\frac{n-1}{2}}\left[\frac{\theta_1\left(\frac{\ep +it}{\pi}+\frac{\sigma}{2\pi}|\frac{2i\ep}{\pi}\right)\theta_1\left(\frac{\ep +it}{\pi}-\frac{\sigma}{2\pi}|\frac{2i\ep}{\pi}\right)\eta\left(\frac{2i\ep}{\pi}\right)^6}{\theta_1\left(\frac{\sigma}{2\pi}|\frac{2i\ep}{\pi}\right)^2\theta_1\left(\frac{\ep +it}{\pi}|\frac{2i\ep}{\pi}\right)^2}\right]^{\frac{2a^2}{n^2}}.  \label{resultnd}
\ea
where we noted that the $a=0$ sector contribution is trivial $\left[\mbox{Tr}(\rho_A)^n\right]_{0}=1$ as the twist operators are identity operator.

Now it is easy to see that the final result (\ref{resultnd}), obtained for the Neumann-Dirichlet boundary condition, is identical to that for the Neumann-Neumann boundary condition computed in \cite{Takayanagi:2010wp}. Thus our pseudo entropy is the same as the entanglement entropy for $|N\lb$.\\

\bibliographystyle{JHEP}
\bibliography{REF}

\end{document}